\documentclass[12pt]{article}

\catcode`\@=11
\@addtoreset{equation}{section}

\global\arraycolsep=2pt
\usepackage{mathrsfs,amsbsy,amssymb,latexsym,amsfonts,amsmath,cite}

\usepackage{graphicx,color}

\newcommand\CL{{\mathcal L}}

\renewcommand\NG{{\mathrm{NG}}}

\newcommand\e{\mathrm{e}}

\newcommand{\adss}[2]{{AdS$_{#1}\times$S$^{#2}$}}

\newcommand{\s}[1]{{S$^{#1}$}}

\newcommand\NR{{\mathrm{NR}}}

\newcommand\Tr{\mathrm{Tr}}
\newcommand\CO{\mathcal{O}}
\newcommand\CW{\mathcal{W}}
\newcommand\nn{\nonumber}

\begin{document}

\begin{flushright}
\parbox{4.2cm}
{OIQP-07-17 \hfill \\ 
NSF-KITP-07-196}
\end{flushright}

\vspace*{0.5cm}

\begin{center}
{\Large \bf
Holography of Non-relativistic String on AdS$_5\times$S$^5$ 
}
\end{center}
\vspace{10mm}

\centerline{\large Makoto Sakaguchi$^{a}$
 and Kentaroh Yoshida$^{b}$
}

\vspace{8mm}

\begin{center}
$^a$ {\it Okayama Institute for Quantum Physics\\
1-9-1 Kyoyama, Okayama 700-0015, Japan} \\
{\tt makoto\_sakaguchi\_at\_pref.okayama.jp}
\vspace{5mm}

$^b$ {\it Kavli Institute for Theoretical Physics, \\ 
University of California,  \\
Santa Barbara,  CA.\ 93106, USA
} 
\\
{\tt kyoshida\_at\_kitp.ucsb.edu}
\end{center}

\vspace{1cm}

\begin{abstract} 
We discuss a holographic dual of a non-relativistic (NR) string on
AdS$_5\times$S$^5$\,. The NR string can be regarded as a semiclassical
string around an AdS$_2$ classical solution corresponding to a straight
Wilson line in the gauge-theory side. The quadratic action with respect
to the fluctuations is composed of free massive and massless scalars,
and free massive fermions on the AdS$_2$ world-sheet. We show that the
complete agreement of the spectra between the NR string and a conformal
quantum mechanics (CQM). Then we show a holographic relation between
normalizable modes of the NR string and wave functions in the CQM. Then
it may be argued from this result that an AdS$_2$/CFT$_1$ would be
realized in AdS$_5$/CFT$_4$\,. We can really discuss a GKPW-type
relation by considering non-normalizable modes of the NR string in
Euclidean signature. Those modes give a source term insertion to the
Wilson line, which can also be regarded as a small deformation of it.
\end{abstract}

\thispagestyle{empty}
\setcounter{page}{0}

\newpage

\section{Introduction}

One of the most important subjects in string theory is AdS/CFT
correspondence \cite{AdS/CFT:M, AdS/CFT:GKPW}, which gives a realization
of holography within the framework of string theory. There is no
rigorous 
proof of it. An important issue toward the proof
is to quantize
type IIB string on \adss{5}{5}. Although it is shown to be classically
integrable \cite{BPR}, its action constructed in \cite{MT} is so
non-linear that it is difficult to quantize it directly. Thus it may be
important to look for a solvable limit.
In fact, the study of pp-wave string \cite{Metsaev,MT2} with Penrose 
limit \cite{Penrose} played an important role in examining AdS/CFT 
even in a non-BPS region \cite{BMN}.

\medskip 

Recently an interesting solvable subsector of AdS$_5$/CFT$_4$ with
non-relativistic (NR) limit\footnote{A non-relativistic limit for a
closed string in flat space was discussed in \cite{KM,GO} earlier than
in AdS$_5\times$S$^5$. } was proposed in \cite{GGK}. After taking the
limit, the full string theory is reduced to a free theory on the AdS$_2$
world-sheet. More precisely speaking, the resulting theory contains
three massive and five massless scalars, and eight massive fermions. It
can also be obtained as a semiclassical limit of the full AdS string
around an AdS$_2$ solution corresponding to a Wilson line \cite{Wilson},
like in the case of pp-wave string \cite{GKP}. The quadratic action of
string and D-branes in the semiclassical limit are computed in
\cite{DGT,SY:NR1,SY:NR2,SY:NR3}.

\medskip 

In this paper we proceed with the previous works
\cite{SY:NR1,SY:NR2,SY:NR3} and discuss a holographic dual of the NR
string. Here we are confined to 
the case of straight Wilson line.
Then the
physical spectrum of the NR string has already been computed in
\cite{ST} and it is used to evaluate the semiclassical partition
function of AdS superstring in \cite{DGT}. Motivated by this spectrum
and the AdS$_2$ world-sheet geometry, it would be worth looking for a
candidate of the dual description of the NR string. The dual theory may
be defined on the Wilson line which is the boundary of the world-sheet
and one-dimensional. The physical normalizable modes of the NR string
decay before reaching the boundary of the AdS$_2$ world-sheet. However,
it may be related to the quantum Hilbert space according to the
Lorentzian AdS/CFT dictionary \cite{BKL}\footnote{For a more
comprehensive discussion on the Lorentzian AdS/CFT see \cite{Marolf}.}.

\medskip 

In fact we can find
that
the spectrum of
 a one-dimensional conformal quantum mechanics (CQM)
\cite{CQM}\footnote{For a recent review of CQM see \cite{review:SQM}.}
completely agrees with that of the NR string under
a certain identification of parameters. The wave functions of the CQM
can be reproduced from the physical modes of the NR string following
\cite{NY}. In addition the two-point function in the CQM can also be
reproduced by evaluating the Wightman function from the bulk NR string.

\medskip

Based on the evidence we may argue an AdS$_2$/CFT$_1$ realized in
AdS$_5$/CFT$_4$ via the so-called ``double holography.'' Eventually, it
would not be so surprising to find the holography of this type, since
the holographic relation of this type has already been observed in the
case of defect CFT \cite{DFO}. It would be helpful for the readers
familiar with the case to imagine replacing AdS$_4\times$S$^2$ with
simply AdS$_2$\,.

\medskip 

Then we can discuss a GKPW-type relation \cite{AdS/CFT:GKPW} between the
NR string and the Wilson loop with a source term insertion by
considering non-normalizable (NN) modes of the NR string. This source
term is given by a one-dimensional integral and sensitive to the
NN modes on the AdS$_2$ string world-sheet. The operators
coupled to the boundary values of the NN modes have
already been clarified in \cite{SY:NR2,SY:NR3}. The insertion of the
source term can be regarded as a deformation of the Wilson
line. Furthermore we discuss supersymmetries preserved under the
insertion of the source term.

\medskip 

This paper is organized as follows.  In section 2 we first introduce the
action of the NR string on AdS$_5\times$S$^5$\,. Then we discuss the
physical spectrum of the NR string.  This is a review of the work
\cite{ST}. In section 3 we find a CQM dual to the NR string. This
section contains some reviews on the basics of CQM.  In section 4 we
first remember a semiclassical derivation of the bosonic part. Then
we discuss the
treatment for the divergent part,
the relation between the two methods: 1) adding a constant $B$-field and
2) the standard method to consider a Legendre transformation.  Next we
discuss a holographic relation, i.e., GKPW-type relation between the NR
string and a straight Wilson line with a source term insertion by
considering NN modes of the NR string. We carefully examine the relation
between the fluctuations and the NN modes, and we show that the
fluctuations do not diverge at the boundary even for the NN
modes. Section 5 is devoted to a conclusion and discussions.

\section{NR string on AdS$_5\times$S$^5$}

We introduce a non-relativistic (NR) string on \adss{5}{5}
\cite{GGK}. In the first place we shall give a brief review of a series
of papers about the action of the NR string on AdS$_5\times$S$^5$\,.
Our convention and notation are also described. Then the physical
spectrum of the NR string obtained in \cite{ST}
is briefly reviewed here.

\subsection{The action of NR string}

The NR limit concerned here is realized by taking the speed of light
transverse to an AdS$_2$ subspace to be infinite\footnote{ More
generally, another NR limit may be considered by making the speed of
light in the directions transverse to an AdS$_p\times$S$^q$ subspace
infinite. In fact, considering 1/2 BPS AdS-branes, NR limits of DBI
actions on AdS$_5\times$S$^5$ can be discussed \cite{SY:NR1}.}, while it
should be kept along the AdS$_2$\,.  The concrete prescription to take
the limit is given in Appendix \ref{nr}.

\medskip 

After taking the NR limit and fixing the $\kappa$-symmetry, the action 
is reduced to the following
form\footnote{ For a heuristic derivation see Appendix \ref{der}.},
\begin{eqnarray}
S^{\rm (NR)} =
-\int\!\!d^2\sigma
\sqrt{-g}\Bigl[
\frac{g^{ij}}{2}\partial_i x^a \partial_j x^a + \frac{1}{R_0^2}x^a x^a 
+\frac{g^{ij}}{2}\partial_i y^{a'}\partial_j y^{a'} 
-2i \bar{\theta}_+ \Gamma^{\mu}\mathbf{v}_{\mu}^i D_{i}\theta_+
\Bigr]\,.  \label{action} 
\end{eqnarray}
Here the string tension has been absorbed by rescaling the variables
$(x^a,y^{a'},\theta_+)$\,. The world-sheet coordinates are
$\sigma^{i}=(\sigma^0,\sigma^1)=(\tau,\sigma)$ and the world-sheet
metric $g_{ij}$ on the world-sheet is
\[
g_{ij}=\eta_{\mu\nu}\mathbf{v}_i^{\mu}\mathbf{v}_j^{\nu} \quad
(\mu,\nu=0,1)\,, \qquad \quad g \equiv \det g_{ij}\,,
\] 
where $\mathbf{v}^{\mu}$ is a zweibein of AdS$_2$\,. That is, the world-sheet
geometry is a two-dimensional AdS space. 
The eight bosonic variables $x^a~(a=1,2,3)$ and $y^{a'}~(a'=1,\ldots,5)$
come from AdS$_5$ transverse to the AdS$_2$ and S$^5$\, respectively. 

\medskip

The fermionic variable $\theta$ is decomposed into the two parts like
$\theta=\theta_+ + \theta_-$ in terms of the two eigenvalues of
$\Gamma_{\ast} \equiv \Gamma_0\Gamma_1\tau_3$ satisfying
$\Gamma_{\ast}^2=1$\,.  That is,
$\Gamma_{\ast}\theta_{\pm}=\pm\theta_{\pm}$\,. The $\kappa$-symmetry is
fixed by taking the condition $\theta_-=0$\,. Note that the action
(\ref{action}) preserves 16 linear supersymmetries and 16 non-linear
supersymmetries. That is, $\mathcal{N}=8$ supersymmetries in two
dimensions are preserved \cite{DGT,GGK}. For an earlier discussion on
the structure of $\mathcal{N}=1$ supermultiplet on AdS$_2$\,, see
\cite{ST}.

\medskip 

The Virasoro condition has already been solved and the action apparently
contains the eight physical components, i.e., three massive and five
massless bosons, and eight massive fermions. The masses of bosons and
fermions are measured by the AdS radius $R_0$\,, and the boson mass is
$m_{\rm B}^2=2/R_0^2$ and the fermion mass is $m_{\rm
F}^2=1/R_0^2$\,. Thus the SO(3)$\times$SO(5) bosonic symmetry is
preserved rather than SO(8)\,. The world-sheet has an SL(2) symmetry.  
Including the fermions the symmetry is
enhanced to OSp(4$^{\ast}|$4).

\medskip 

In general, unphysical components may be contained implicitly through
the AdS$_2$ zweibein $\mathbf{v}^{\mu}$\,. Those can however be removed
by taking a static gauge \cite{GGK} and the zweibein does not depend on 
the unphysical components. Hereafter we will 
assume that the static gauge is taken.
According to the gauge-fixing to the static gauge, in working on
Euclidean AdS$_5\times$S$^5$, the world-sheet AdS$_2$ should also be
Euclidean AdS$_2$ (EAdS$_2$)\,.

\subsection{Physical spectrum of NR string}

We will discuss the quantization of normalizable solutions of a free
scalar field $\phi$ on AdS$_2$ in Lorentzian signature. It should be
regarded as one of the scalar components contained in the action of the
NR string. Namely, $\phi =x^a~{\rm or}~y^{a'}$\,.  Here we restrict
ourselves to the bosonic part, but the argument for the fermion is also
given in \cite{ST}.

\medskip 

The world-sheet AdS$_2$ geometry is described by the metric:
\begin{eqnarray}
ds^2 =g_{ij}dx^{i}dx^{j}= \frac{R_0^2}{\cos^2 \rho}(-dt^2 + d\rho^2)\,.
\label{ads2}
\end{eqnarray}
Here the time $t$ is the global time and $-\pi/2\leq \rho \leq
\pi/2$\,. The boundary has the topology $\mathbb{R}\times S^0$ and so
there are two time-like boundaries at $\rho=\pm\pi/2$\,.

\medskip 

The classical equation of
motion is given by
\begin{equation}
\cos^2\rho\left(-\partial_{t}^2 +\partial_{\rho}^2\right)\phi
-m^2\phi =0\,, \label{eq}
\end{equation}
where $m^2=R_0^2m_B^2=2$ for $x^a$ and $m^2=0$ for $y^{a'}$.
The formal solution is typically written by the Gegenbauer
polynomial\footnote{For some properties of Gegenbauer polynomials, see
Appendix \ref{gegen}.} $C_{\alpha}^{\lambda}(z)$ \cite{ST,NY}: 
\begin{eqnarray}
\phi^{\lambda,\pm}_{\omega} (t,\rho) = {\rm e}^{\pm i \omega t} 
(\cos\rho)^{\lambda} C_{\omega-\lambda}^{\lambda}(\sin\rho)\,. \label{sol}
\end{eqnarray}
In order for (\ref{sol}) to be the solution of (\ref{eq})\,, 
$\omega$ has to take the discrete value as 
\[
 \omega  = n + \lambda \qquad (n=0,1,\cdots )\,,
\] 
and furthermore $\lambda$ takes $\Delta$ or $1-\Delta$\,, where $\Delta$
is defined as
\begin{eqnarray}
\Delta \equiv \frac{1}{2}\left(1+\sqrt{1+4m^2}\right)\,. \nn
\end{eqnarray}

Then in order to discuss the normalizability we have to introduce the
Klein-Gordon norm
\begin{equation}
(\phi_1,\phi_2) \equiv i\int\!d\rho\,\sqrt{-g}g^{tt}\left[
\phi_1\partial_t \bar{\phi}_2 - \bar{\phi}_2^{\ast}\partial_t \phi_1
\right]\,. \label{ip}
\end{equation}
The solution (\ref{sol}) is normalizable with (\ref{ip}) 
if and only if $\lambda$ is real and $\lambda > -1/2$\,. 
In particular, the reality of $\lambda$ is equivalent to the  
Breitenlohner-Freedman (BF) bound \cite{BF}: 
\begin{eqnarray}
m^2 \geq -\frac{1}{4}\,. \label{BF}
\end{eqnarray}
Furthermore by imposing
the unitarity $\lambda$ can be further restricted and the additional
condition is $\lambda >0$\,. 
Thus we will take $\lambda$ as $\Delta$\,. 

\medskip 

The quantization can be performed by following the standard procedure. 
First of all, the field $\phi(t,\rho)$ is expanded as 
\begin{eqnarray}
\phi^{\Delta}(t,\rho) = \sum_{n=0}^{\infty}a_n \phi_{n}^{\Delta}(t,\rho) 
+ \sum_{n=0}^{\infty}a_n^{\dagger}\bar{\phi}_n^{\Delta}(t,\rho)\,,
\end{eqnarray}
where the $n$-th mode is defined as 
\begin{eqnarray}
\phi_n^{\Delta}(t,\rho) \equiv
 c(\Delta)\sqrt{\frac{n!}{\Gamma(n+2\Delta)}}
{\rm e}^{-i(n+\Delta)(t+\pi/2)}(\cos\rho)^{\Delta}
C_n^{\Delta}(\sin\rho)\,.
\label{n}
\end{eqnarray}
The normalization constant $c(\Delta)$ is defined as 
\[
 c(\Delta) \equiv \frac{\Gamma(\Delta)2^{\Delta -1}}{\sqrt{\pi}}\,,
\]
and it has been fixed by the following conditions:
\begin{eqnarray}
(\phi_m^{\Delta},\phi_n^{\Delta})=\delta_{m,n}\,, \quad 
(\bar{\phi}_m^{\Delta},\bar{\phi}_n^{\Delta}) = - \delta_{m,n}\,, \quad 
(\phi_{m}^{\Delta},\bar{\phi}_n^{\Delta}) =0\,. \nn
\end{eqnarray}
Note that the normalizable modes are decaying as approaching the
boundary, as easily shown from the behavior of the
Gegenbauer polynomials at the boundary:
\[
 C_n^{\Delta}(1) = \frac{\Gamma(n+2\Delta)}{\Gamma(2\Delta)n!}\,. 
\]

\medskip 

We can canonically quantize $\phi(t,\rho)$ and the creation and
annihilation operators $a^{\dagger}_n$ and $a_m$ satisfy the commutation
relations:
\begin{eqnarray}
[a_m,a_n^{\dagger}]=\delta_{m,n}\,, \quad 
[a_m,a_n]=[a_m^{\dagger},a_n^{\dagger}] = 0\,. \nn 
\end{eqnarray} 
Then the Fock vacuum is defined as 
\[
 a_n|0\rangle = 0 \qquad (\forall n=0,1,\cdots)\,,
\]
and the Fock space $\mathcal{F}^{\Delta}$ is spanned as 
\[
 \mathcal{F}^{\Delta} =
 \bigoplus_{n_1,\ldots,n_k}\mathbb{C}\,a_{n_1}^{\dagger}
\cdots a_{n_k}^{\dagger}|0\rangle\,. 
\]
The normalizable modes on AdS$_2$ in the coordinates (\ref{ads2}) have
been studied well in \cite{ST} with the help of supersymmetries. By using
them an effective potential of a free scalar field on AdS$_2$ has been
computed in \cite{IO}. The technique to compute the effective potential
has been applied to computing the one-loop vacuum energy of the
semiclassical action of AdS superstring in \cite{DGT}. 
Note that the Poincare disk should be considered rather than the strip 
of (\ref{ads2}) if the circular case is considered. 

\medskip 

Here the time coordinate $t$ in the metric (\ref{ads2}) is the global
time and the time translation invariance is associated with the global
AdS energy $E$\,. 
The energy $E$ is given as the eigen-value of the Hamiltonian
represented by
\begin{eqnarray}
H = \sum_{n=0}^{\infty}(n+\Delta)a_n^{\dagger}a_n\,,
\end{eqnarray}
up to the zero point energy\footnote{The zero-point energy is not important
because it must be canceled out together with the contributions from
the other degrees of freedom on the AdS$_2$ due to the
supersymmetries.}. 
Then the energy of the $n$-th mode is
given by 
\begin{eqnarray}
 E_n = n + \Delta\,. \label{ene} 
\end{eqnarray}

In the next section we will find a dual CQM whose spectrum completely
agrees with (\ref{ene}).

\section{The dual of the NR string} 

We discuss a CQM, which is expected to be
dual to the quadratic fluctuations around the AdS$_2$\,. Most of the
results here have already been given in \cite{CQM} (For a recent review
see \cite{review:SQM}). We will give an interpretation of the results
in the context of AdS/CFT duality.  In particular, the wave functions of
the CQM coincide with the quantum states of the scalar fields on AdS$_2$
\cite{ST}.

\subsection{CQM and its algebra}

The Hamiltonian of CQM is given by \cite{CQM}
\begin{eqnarray}
H = \frac{1}{2}p^2 + \frac{g}{2x^2}\,, \label{hamiltonian}
\end{eqnarray}
where $g$ is a coupling constant. This model is often called DFF model. 

\medskip 

The one-dimensional conformal group is SO(2,1)=SL$(2,\mathbb{R})$ and its Lie
algebra contains three generators, $H$: Hamiltonian, $D$: dilatation,
and $K$: special conformal. The generators $D$ and $K$ are given by,
respectively,
\begin{eqnarray}
D = -\frac{1}{4}\left(px+xp\right)\,, \qquad K = \frac{1}{2}x^2\,.\nn
\end{eqnarray}
These generators obey the following relations:
\begin{eqnarray}
[H,D]=iH\,, \quad [K,D]=-iK\,, \quad [H,K] = 2iD\,.\nn
\end{eqnarray}

\begin{figure}
 \begin{center}
  \includegraphics[scale=.5]{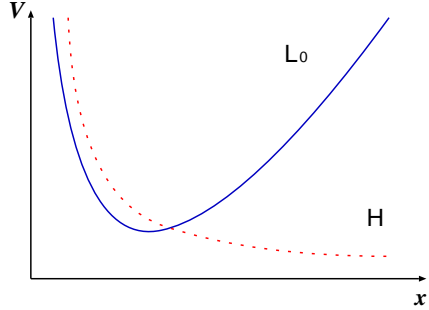}
 \end{center}
\vspace*{-0.5cm}
\caption{\footnotesize The potentials for $H$ and $L_0$\,.}
\label{pot:fig}
\end{figure}

It is awkward to compute the energy spectrum of $H$ because the
potential has no minimum and the spectrum is continuous. Hence according
to \cite{CQM} let us change the basis of the algebra 
from $(H,D,K)$ to $(R,D,S)$\,, where $R$ and $S$ are defined as,
respectively, 
\begin{eqnarray}
 R \equiv \frac{1}{2}\left(a H + \frac{1}{a}K\right)\,, \qquad 
 S \equiv \frac{1}{2}\left(-aH + \frac{1}{a}K\right)\,. \label{alg}
\end{eqnarray} 
Here $a$ is a constant parameter with dimension of length-squared and it
can be understood as a mass parameter of the theory described by
$1/\sqrt{a}$\,. When we regard $R$ as another Hamiltonian, the potential
has a minimum and its spectrum becomes discrete. For the shapes of the
potentials see Fig.\ \ref{pot:fig}.

Then the new algebra is given by 
\begin{eqnarray}
[D,R]=iS\,, \quad [S,R]=-iD\,, \quad [S,D]=-iR\,. \nn 
\end{eqnarray}
This is nothing but the SO(2,1) algebra and the generator $R$
corresponds to a compact rotation U(1) $\subset$ SO(2,1). The others
describe hyperbolic non-compact rotations. In the context of AdS/CFT,
the one-dimensional conformal group SO(2,1) is a subset of the
four-dimensional conformal group SO(2,4)\,. Then the compact rotation
$R$ corresponds to the time translation symmetry with respect to the
global AdS time, which is generated by $H_{\rm full}=\frac{1}{2}(P_0 +
K_0)$~($P_0$ an $K_0$ are zeroth components of translation and special
conformal generators of the SO(2,4))\,. This fact plays an important
role when considering the correspondence
between the spectra of the CQM and a scalar field on AdS$_2$\,.

\subsection{Discrete eigen-value problem}

Next let us consider the energy eigen-value problem: 
\[
 R \beta_n = E_n \beta_n \qquad (n=0,1,\ldots,)\,.  
\]
The energy for the normalizable ground state, $E_0$\,, is
 given by 
\begin{eqnarray}
E_0 = \frac{1}{2} + \frac{1}{2}\sqrt{g + \frac{1}{4}}\,, \nn 
\end{eqnarray}
and the energy eigen-value $E_n$ is
\begin{eqnarray}
E_n = E_0 + n\,.  \label{ene-cqm}
\end{eqnarray}
Note that (\ref{ene-cqm}) completely agrees with the quantum spectrum \eqref{ene} of
a scalar field on AdS$_2$ \cite{ST}, if we identify the scalar mass with
the coupling constant $g$ as follows:
\begin{eqnarray}
 g = 4 m^2 + \frac{3}{4}\,. \label{iden}
\end{eqnarray}  
Then $E_0$ is nothing but $\Delta$\,, i.e., 
\[
E_0 = \Delta\,. 
\]
The parameter $a$ appeared in (\ref{alg}) should be identified with the
square of the AdS radius $R_0^2$\,. The energy eigen-states
$\{\beta_n\}_{n=0,\cdots}$ correspond to those of the scalar field on
AdS$_2$\,, namely normalizable modes. In fact, the set of $\beta_n$ can
be reconstructed from quantized scalar fields on the bulk AdS$_2$\,, by
following \cite{NY}. This recipe will be available in the next
subsection. Thus the result realizes the claim of \cite{BKL}. 

\medskip 

According to (\ref{iden}), the massive (massless) scalars with
$m^2=2~(m^2=0)$ correspond to the CQMs with the following coupling
constant:
\begin{eqnarray}
 g_{\rm AdS}=\frac{35}{4} \quad (m^2=2,~\Delta=2)\,, 
\qquad g_{\rm S}=\frac{3}{4} \quad (m^2=0,~\Delta =1)\,.  \label{want}
\end{eqnarray}

\medskip 

Here we should note the relation between the coordinate system and the
basis of the conformal algebra. The Hamiltonian (\ref{hamiltonian}) is
associated with the time translation symmetry in the Poincare time, and
$R$ is with the Cartan generator of SO(2,1), which describes the
compact rotation. Thus, in the context of AdS/CFT duality, the
Hamiltonian (\ref{hamiltonian}) corresponds to the Poincare energy and
the $R$ to the global AdS energy.

\medskip 

An interesting question is whether the CQM with the coupling
(\ref{want}) can be derived directly from $\mathcal{N}$=4 SYM. This is
an open problem to be investigated in the future, and now we have no
answer to this question.

\medskip 

Finally we shall give some comments on the above CQM argument below.

\subsubsection*{Comment on the bound for the coupling}

It has been known that the following bound
\begin{eqnarray}
 g \geq -1/4 \nn
\end{eqnarray}
should be satisfied so that the energy spectrum is bounded from below.
This statement is obtained from the study of the spectrum of the system
(\ref{hamiltonian})\,. From the viewpoint of AdS/CFT duality we can
easily understand this result as the  BF bound
for the AdS$_2$ case (\ref{BF}). This is surely consistent with the
identification (\ref{iden})\,.

\subsubsection*{Conformal symmetry}

By redefining the dilatation operator\footnote{The definition of
dilatation in \cite{review:SQM} is different from the one in \cite{CQM}.
The dilatation is $D$ in \cite{CQM} and $\tilde{D}$ in
\cite{review:SQM}.} as $\tilde{D} = -2 D$ and introducing the following
linear combinations
\begin{eqnarray}
L_0 = \frac{1}{2}\left(aH+\frac{K}{a}\right)\,, \qquad 
L_{\pm 1} = \frac{1}{2}\left(aH-\frac{K}{a}\mp i \tilde{D}\right)\,,
\label{stro-alg}
\end{eqnarray}
we can find the SL(2,$\mathbb{R}$) algebra in the Virasoro form,
\begin{eqnarray}
[L_1,L_{-1}]=2L_0\,, \qquad [L_0,L_{\pm 1}]=\mp L_{\pm 1}\,. 
\nn 
\end{eqnarray}
Here $L_0$ is the same as $R$\,, i.e., $L_0 = R$\,. Then the eigen-value
of $L_0~(R)$ gives the scaling dimension of the field and hence the
energy eigen-value is nothing but the scaling dimension. The ground
state in terms of $R$ gives a primary field, and the excited states are
secondary fields because $L_{-1}$ gives the excited states. Eventually
this fact matches the argument for the spectrum of a scalar field on
AdS$_2$ \cite{strominger,NY}\footnote{See also \cite{JY}
for an earlier work on a two-dimensional black hole/CQM.}.

\subsubsection*{Supersymmetric CQM}

Finally we shall comment on the supersymmetric case. Including the
fermionic degrees of freedom, the bosonic symmetry
SL(2)$\times$SO(3)$\times$SO(5) is enhanced to the supergroup
OSp$(4^\ast|4)$\,. The action of the supersymmetric CQM related to the
supergroup is discussed in \cite{osp}. The action in \cite{osp} would be
helpful for our purpose.

\subsection{Wave functions from normalizable modes on AdS$_2$}

Our argument can be further confirmed by computing a two-point function
in the CQM from the degrees of freedom in the bulk AdS$_2$\,, i.e., by
extracting the wave function from (\ref{n})\,. The recipe has already
been given in \cite{NY}. The variable in the dual CQM can be defined
from (\ref{n}) as follows:
\begin{eqnarray}
\varphi^{\Delta}(t) \equiv \lim_{\rho\to\pi/2}
\frac{\Gamma(2\Delta)}{c(\Delta)}(\cos\rho)^{-\Delta}
\phi^{\Delta}(t,\rho)\,. \nn 
\end{eqnarray}
In terms of the modes, we can find that 
\begin{eqnarray}
\beta_n^{\Delta}(t) \equiv \lim_{\rho\to\pi/2}
\frac{\Gamma(2\Delta)}{c(\Delta)}(\cos\rho)^{-\Delta}
\phi_n^{\Delta}(t,\rho)
= \sqrt{\frac{\Gamma(n+2\Delta)}{n!}}\,{\rm e}^{-i(n+\Delta)(t+\pi/2)}
\,. \nn 
\end{eqnarray}
Then the two-point function can be computed from the boundary behavior
of the Wightman function
\[
\langle 0|\phi^{\Delta}(t_1,\rho_1)\phi^{\Delta}(t_2,\rho_2) |0\rangle
\] 
on the bulk AdS$_2$\,. 

\medskip 

The two-point function
$\langle\varphi^{\Delta}(t_1)\varphi^{\Delta}(t_2)\rangle$ is evaluated as
follows: 
\begin{eqnarray}
\langle\varphi^{\Delta}(t_1)\varphi^{\Delta}(t_2)\rangle &=& 
\lim_{\rho_1,\rho_2\to\pi/2}\left(\frac{\Gamma(2\Delta)}{c(\Delta)}\right)^2
(\cos\rho)^{-2\Delta} \langle\phi^{\Delta}(t_1,\rho_1)
\phi^{\Delta}(t_2,\rho_2)\rangle \nn \\ 
&=& \sum_{n= 0}^{\infty}\beta_n(t_1)\bar{\beta}_n(t_2) 
= \frac{\Gamma(2\Delta)
{\rm e}^{i \Delta(t_1+t_2)}}{({\rm e}^{it_1}
-{\rm e}^{it_2})^{2\Delta}}\,. \label{propag}
\end{eqnarray}
In the above computation the following formulae have been utilized:
\[
 \sum_{n=0}^{\infty}\frac{\Gamma(n+2\Delta)}{n!}x^n 
= \frac{\Gamma(2\Delta)}{(1-x)^{2\Delta}}\,, \qquad \Gamma(n+1) = n!\,. 
\] 

\medskip 

The expression of (\ref{propag}) is written in Lorentzian signature and
it needs some algebra to find a familiar form. Let us first perform the
Wick rotation: $t=-it_{\rm E}$\,. Then EAdS$_2$ with Poincare metric,
\[ds^2= \frac{d\tau_{\rm P}^2 +dz^2}{x^2}
\]
is related to the Lorentzian metric (\ref{ads2}) via the following coordinate 
transformation:
\[
\tau_{\rm P} = {\rm e}^{t_{\rm E}}\sin\rho\,, 
\qquad z={\rm e}^{t_{\rm E}}\cos\rho\,.
\]
As approaching the boundary $z\rightarrow 0$\,, the Poincare time
behaves as $\tau_{\rm P} \to {\rm e}^{t_{\rm E}}$\,. The conformal field
$\varphi^{\Delta}(t)$ has the conformal weight $\Delta$ and transforms
to $\hat{\varphi}(x)$ according to 
\[
\varphi^{\Delta}(t)(dt)^{\Delta} 
= \hat{\varphi}^{\Delta}(\tau_{\rm P})(d\tau_{\rm P})^{\Delta}\,.
\] 
Then we can obtain the correlation function 
\begin{eqnarray}
\langle \hat{\varphi}^{\Delta}(\tau_{P1})  
\hat{\varphi}^{\Delta}(\tau_{P2})\rangle 
= \frac{\Gamma (2\Delta)}{(\tau_{\rm P1}-\tau_{\rm P2})^{2\Delta}}\,.\nn
\end{eqnarray}
This is nothing but the two-point function of the corresponding CQM
\cite{CQM}.

\medskip 

From the above argument it is natural to look for the GKPW-type relation
\cite{AdS/CFT:GKPW} as the next step. Then we have to consider NN modes,
which are relevant to the operator insertions in the boundary theory.
It will be discussed in detail in the next section.

\section{Holography of NR string and Wilson line}
\label{3}

In this section we shall discuss a GKPW-type relation
\cite{AdS/CFT:GKPW} between the NR string and a Wilson line with local
operator insertions, by considering NN modes of the NR string. We will
work in Euclidean signature here.

\medskip 

In the first place we remember that the NR string action (\ref{action})
can also be obtained as a semiclassical approximation around an AdS$_2$
solution corresponding to a Wilson loop \cite{DGT,SY:NR2,SY:NR3}.  Then
we have to be careful for a field redefinition of the variables.  This
is sensitive to the order of the divergence of NN modes of
(\ref{action}) near at the boundary. Finally we propose a GKPW-type
relation and clarify the related operator insertion in the gauge-theory
side.

\subsection{NR string from semiclassical limit} 

Let us remember the semiclassical approximation of the full AdS
superstring around a static AdS$_2$ solution. Note that the AdS$_2$
classical solution satisfies the equations of motion obtained from the
full action. Here we are confined to the bosonic part for simplicity and
take the Nambu-Goto (NG) formulation with a static gauge.

\medskip

The bosonic NG action is given by
\[
 S_{\rm NG} = \frac{\sqrt{\lambda}}{2\pi}\int\!d^2\sigma\,\sqrt{\det g}\,, 
\qquad g_{ij} = \partial_i X^{M} \partial_j X^N G_{MN}
\]
in the Euclidean Poincare coordinates
\[
 ds^2=G_{MN}dX^M dX^N = \frac{1}{z^2}\left(dX^{m}dX^m +
 dz^2\right) + d\Omega_5^2\,. 
\]
Here the AdS radius $R_0$ is already absorbed into the definition of 't
Hooft coupling $\lambda \equiv
Ng_\mathrm{YM}^2= R_0^4/\alpha'{}^2$\,.

\medskip 

Next the action can be expanded around the static classical solution as
follows:
\[
X^0=\tau\,, \quad z=\sigma\,, \quad X^a =
0+ \sqrt{2\pi}{\lambda^{-1/4}}\tilde{x}^a\,, \quad 
Y^{a'} = 0 + \sqrt{2\pi}\lambda^{-1/4}\tilde{y}^{a'}\,,
\]
where $Y^{a'}$ is the tangent coordinate on \s{5}.
Then the induced metric can be expanded as 
\begin{eqnarray}
&&  g_{ij} = g_{0ij} + 2\pi \lambda^{-1/2}\left[\frac{1}{\sigma^2}\partial_i
\tilde x^a \partial_j\tilde x^a 
+ \partial_i \tilde y^{a'}\partial_j\tilde y^{a'}\right]
+\cdots \,, 
\nonumber \\ 
&& g_{0ij} = \frac{1}{\sigma^2}\delta_{ij}\,. \nonumber 
\end{eqnarray} 
The resulting action is given by 
\begin{eqnarray}
S_{\rm NG} &=& S_{(0)} + S_{(2)} +\mathcal{O}(\lambda^{-1/4})\,, \nonumber \\ 
S _{(0)} &=& \frac{\sqrt{\lambda}}{2\pi}\int\! d^2\sigma\,\sqrt{\det g_0} 
= \frac{\sqrt{\lambda}}{2\pi}\int\! d\tau\,\frac{1}{\epsilon}\,, 
\label{S0} \\
S_{(2)} &=& \frac{1}{2}\int\! d^2\sigma\,\sqrt{\det g_0} 
g_0^{ij}\left(\frac{1}{\sigma^2}
\partial_i \tilde{x}^a\partial_j \tilde{x}^a 
+ \partial_i \tilde y^{a'}\partial_j \tilde y^{a'} \right)
\,.  
\label{S2 in fluc}
\end{eqnarray}
Here $S_{(i)}~(i=0,1,2,\ldots)$ denote the action with the $i$-th order
quantum fluctuations.
For the
evaluation of $S_{(0)}$ the cut off $\epsilon$ has been introduced. With
the equations of motion, the first-order contribution should vanish,
i.e., $S_{(1)}=0$\,, and hence it will not be touched below.

\medskip 

The zero-th order part gives a volume factor and it diverges. We can
treat well this divergence by adding a boundary term.
It will be discussed in the next subsection. As a result,
for the classical solution concerned here, $S_{(0)}$ should vanish.

\medskip 

When the 't Hooft coupling $\lambda$ is taken to be sufficiently large
and the fluctuations are not divergent, the higher-order terms can be
ignored. Then the approximation keeping the leading terms with lower
order is valid. That is, the leading contribution is the quadratic
fluctuations described by $S_{(2)}$\,.

\medskip 

By performing a field redefinition,
\begin{eqnarray}
{x}^a=\frac{1}{\sigma} \tilde {x}^a\,,~~~
y^{a'}=\tilde y^{a'}\,, \label{redefinition}
\end{eqnarray}
the second order action can be rewritten as 
\begin{eqnarray}
 S_{(2)} &=& S_\NR
 + \frac{1}{2}\int\! d^2\sigma\,\partial_{\sigma}
\left(\frac{1}{\sigma} {x}^a {x}^a\right)\,
\label{S2 in xy}
\end{eqnarray}
where
\begin{eqnarray}
S_\NR&=& \frac{1}{2}\int\! d^2\sigma\,\sqrt{\det g_0} \left[
g_0^{ij}\left(
\partial_i {x}^a\partial_j {x}^a 
+ \partial_i y^{a'}\partial_j y^{a'} \right) + 2  {x}^a {x}^a
\right] \,.
\label{NR action}
\end{eqnarray}
Thus the action $S_{(2)}$ is nothing but the bosonic action $S_\NR$ of
the NR string discussed in the previous subsection, up to the last,
surface term.  The existence of this surface term will play an important
role in our argument later in section 4.3.

\medskip 

The computation here will be available to estimate the magnitude of the
fluctuations around the classical solution when we consider
NN modes later.

\subsection{Notes on boundary conditions}

As is well known, there should be a divergence in the classical action for
a classical sting solution corresponding to a Wilson loop
\cite{Wilson}. Hence a regularization is necessary and then an
appropriate boundary condition has to be imposed to remove the cut-off
dependence \cite{bound}.

\medskip

Here it would possibly be interesting to see a different method to
remove the divergence. This is to introduce a coupling to a constant
NS-NS
$B$-field into the string action \cite{GGK}. This method has been
utilized instead of the standard techniques \cite{bound}. 
Hereafter we will argue that the two methods may be equivalent.  

\medskip 

Let us first remember the standard method given in \cite{bound}. It is
easy to see that
\begin{eqnarray*}
\delta S_\NG=\int d\tau P_M^\sigma \delta X^M |_{\sigma=0}\,, \qquad 
P_M^\sigma=\frac{\partial\CL}{\partial(\partial_\sigma X^M)}\,, 
\end{eqnarray*}
where the equations of motion have been used. It implies that the action
$S_\NG$ is a functional of $X^M$ at the boundary.  On the other hand,
the Wilson loop is a functional of $X^m$ and $\dot Y^{m'}$\,, where
$Y^{m'}$ are Cartesian coordinates on $R^6$:
\[
(dY^{m'})^2=dz^2+z^2d\Omega_5^2\,.
\]  
Hence it leads us to consider the Legendre transformation
\begin{eqnarray}
S&=&S_\NG+S_L\,, \qquad 
S_L=-\int d\tau P^\sigma_{m'}Y^{m'}
|_{\sigma=0}\,,
\end{eqnarray}
and then $S$ is a functional of $X^m$ and $\dot Y^{m'}$ at the boundary.
The $S_L$ is evaluated in the static gauge as
\begin{eqnarray}
S_L=
-\frac{\sqrt{\lambda}}{2\pi}\int d\tau \frac{1}{2}
\frac{\partial_\sigma z^2}{z^2}\Big|_{\sigma=0}
=-\frac{\sqrt{\lambda}}{2\pi}\int d\tau \frac{1}{\epsilon}\,.
\end{eqnarray}
This term cancels out $S_{(0)}$ in \eqref{S0} as expected. 

\medskip 

Next we consider another method to introduce a constant $B$-field \cite{GGK}.
It gives $H=dB =0$ and does not change the AdS$_5\times$S$^5$ 
background. 
Thus we may include a coupling to the $B$-field
\begin{eqnarray}
S_B&=&\frac{1}{2\pi\alpha'}\int {}^{\ast}B
=\frac{1}{2\pi\alpha'}\int d^2\sigma \epsilon^{ij}
\partial_iX^M\partial_jX^NB_{MN}\,. \label{4.8}
\end{eqnarray}
Let us consider the following closed two-form
\begin{eqnarray}
B=-R_0^2E^0\wedge E^z
=-\frac{R^2_0}{z^2}dX^0\wedge dz\,,
\end{eqnarray}
where $E$ is a vielbein\footnote{Here one may suspect here that a
special $B$-field is taken by hand. But it would be fixed from the BPS
condition for the classical solution. In fact, it could be done in the
case of flat space \cite{GKT}. It would be nice to consider the same
analysis for the AdS case.}. Then (\ref{4.8}) can be rewritten as
\begin{eqnarray}
S_B=
\frac{\sqrt{\lambda}}{2\pi}\int d\tau  \frac{1}{\sigma}
\Big|_{\sigma=0}^{\sigma=\infty}
=-\frac{\sqrt{\lambda}}{2\pi}\int d\tau \frac{1}{\epsilon}\,. 
\end{eqnarray}
This cancels out $S_{(0)}$ in \eqref{S0}.

\medskip 

Thus we can conclude that the two methods work well to cancel the
divergence coming from $S_{(0)}$\,. In both cases we have treated the
additional terms, $S_L$ and $S_B$, as classical contributions.  This is
because these terms are added as background configurations which do not
fluctuate. Therefore the fluctuations originate only from $S_\NG$.

\medskip

Furthermore the Legendre transformation of $S_\NG+S_B$ may be considered
by adding
\[
S_L=-\int d\tau P_z^\sigma z |_{\sigma=0}\,, \qquad 
P_z^\sigma\equiv \frac{\partial(\CL_\NG+\CL_B)}{\partial(\partial_\sigma
z)}\,.
\]
Then it can be easily shown that $P_z^\sigma=0$ and $S_L$ should
vanish. That is, the redundant divergent term does not arise.

\medskip

In the next section \ref{nn}, we evaluate the classical value of the
quadratic action, $S_{(2)}[\Phi_0]$ by substituting the NN solution into
the $S_{(2)}[\Phi]$\,.  This should not be confused with $S_{(0)}$\,,
which now corresponds to a Wilson loop.  Finally, in section \ref{hol},
we discuss a holographic interpretation of $S_{(2)}[\Phi_0]$\,.

\subsection{NN modes on AdS$_2$}
\label{nn}

In the Euclidean case there exists NN solution only, 
while in the Lorentzian case normalizable solutions should be considered as well as NN modes. Hence
the Lorentzian case is more complicated than the Euclidean case. The
advantage in Euclidean signature was advocated in \cite{AdS/CFT:GKPW}.

\medskip

From now on we shall discuss NN solutions of the classical
equation of motion derived from $S_{(2)}$\,.  In Euclidean signature the
world-sheet metric of EAdS$_2$ is given by
\begin{eqnarray}
ds^2 = \frac{1}{\sigma^2}(d\tau^2 + d\sigma^2)\,. \nn 
\end{eqnarray}
This is a two-dimensional Poincare metric and the boundary is at
$\sigma=0$\,.  

\medskip 

The NN solution $\Phi^I(\sigma,\tau)$ is specified by the behavior near
the boundary
\begin{eqnarray}
\Phi^I (\sigma,\tau) \rightarrow \sigma^{1-\Delta}\Phi_0^I(\tau) 
\qquad (\sigma\to 0)\,,
\label{NN boundary behavior}
\end{eqnarray}
where the index $I$ describes the eight transverse directions 
and $\Phi^I=(x^{a},y^{a'})$\,. The $\Delta$ is fixed through 
the mass $m_{\rm B}$ of the variable $\Phi^I$ by 
\[
\Delta = \frac{1}{2}\left(1+\sqrt{1+4R_0^2m_{\rm B}^2}\right)\,,
\]
and it implicitly depends on the index $I$\,. From this behavior we find
that $x^a$ diverges near the boundary since $\Delta=2$ for $x^a$\,. Then
one might think that the divergence would break the semiclassical
approximation around the AdS$_2$ solution.

\medskip 

But it is worth noting that the divergence of $x^a$ does not imply that
the fluctuation around the AdS$_2$ solution $\tilde x^a$ should also
diverge. This is mainly because $x^a$ and $\tilde x^a$ are related by
\eqref{redefinition}: $x^a=\frac{1}{\sigma}\tilde x^a$\,. Actually,
$\tilde x^a$ does not diverge as we will see below, and 
the semiclassical approximation may still be valid even in the
presence of the NN modes. 

\medskip 

The NN mode $\Phi^I$ is completely determined from the
boundary value $\Phi_0^I(\tau)$ as follows:
\begin{eqnarray}
\Phi^I (\sigma,\tau) = \int\!d\tau'\,
K_{\Delta}(\sigma,\tau;\tau')\Phi_0^I(\tau')\,. \label{nn-mode}
\end{eqnarray}
Here $K_{\Delta}(\sigma,\tau;\tau')$ is the bulk-to-boundary propagator 
\cite{KW,prop} defined as 
\begin{eqnarray}
K_{\Delta}(\sigma,\tau;\tau') \equiv 
\pi^{-1/2}\frac{\Gamma (\Delta)}{\Gamma \left(\Delta -\frac{1}{2}\right)}
\frac{\sigma^{\Delta}}{(\sigma^2+(\tau-\tau')^2)^{\Delta}}\,.
\label{prop}
\end{eqnarray}

\medskip 

Let us consider the surface term in $S_{(2)}$ given in \eqref{S2 in xy}.
Its classical value is
\begin{eqnarray}
-\frac{1}{2}\int d\tau\frac{1}{\epsilon^3}(x_0^a)^2\,, 
\label{sft}
\end{eqnarray}
where $x_0^a$ is defined from the asymptotic behavior around the
boundary 
\begin{eqnarray}
x^a \to \frac{1}{\epsilon}x^a_0 \qquad (\sigma \to \epsilon)\,. \nonumber
\end{eqnarray}
On the other hand, by
integrating by part 
$S_\NR$ in \eqref{NR action} can be rewritten as 
\begin{eqnarray}
S_\NR&=&
-\frac{1}{2}\int\! d^2\sigma\, \left[
x^a\partial^2 x^a+y^{a'}\partial^2y^{a'}
+\frac{2}{\sigma^2}x^ax^a \right]\nonumber \\ 
&& \qquad  +\frac{1}{2}\int\! d^2\sigma\, \partial_\sigma\left(
x^a\partial_\sigma x^a+y^{a'}\partial_\sigma y^{a'}
\right)
\label{S_NR partial}
\end{eqnarray}
The contribution of the surface term is evaluated as
\[
\frac{1}{2}\int\!d\tau\,\frac{1}{\epsilon^3}x_0^2\,,
\]
and it cancels out (\ref{sft})\,. Thus $S_{(2)}$ is equivalent to the
first line in \eqref{S_NR partial}. That is, 
\begin{eqnarray}
S_{(2)} = -\frac{1}{2}\int\! d^2\sigma\, \left[
x^a\partial^2 x^a+y^{a'}\partial^2y^{a'}
+\frac{2}{\sigma^2}x^ax^a \right]\,. \label{s2}
\end{eqnarray}
The convergence of the $\sigma$-integration in  
\eqref{s2}
 gives the precise unitarity bound on the dimension of a scalar
\cite{KW}.
Hence $S_{(2)}$ may be regarded as
the precise action of the NR string including the boundary term to
ensure the cancellation of the divergence at the boundary.

\medskip 

In fact, $S_{(2)}$ in \eqref{S2 in fluc} is nothing but the proposed
action to define the two-point function \cite{KW}\footnote{As noted
there, for $\Delta\ge 3/2$, there still remain boundary divergences.
They may correspond to contact terms in the Wilson loop expansion
including $\delta x$ in Appendix D.}.  Then the classical value of
$S_{(2)}$ with $\tilde \Phi^I \equiv (\tilde x^a, \tilde y^{a'})$ is
given by
\begin{eqnarray}
S_{(2)}&=& - \sum_{I=1}^8
\lim_{\sigma\to 0}\sigma^{2-2\Delta}\frac{1}{2}\int\! d\tau\, 
\tilde \Phi^I \partial_\sigma\tilde  \Phi^I 
\,. 
\end{eqnarray}
By using (\ref{nn-mode}), the resulting value of $S_{(2)}$ 
is evaluated as \cite{KW}
\begin{eqnarray}
S_{(2)}[\Phi_0] = -\sum_{I=1}^8\left(\Delta-\frac{1}{2}\right)\pi^{-1/2}
\frac{\Gamma (\Delta)}{\Gamma \left(\Delta-\frac{1}{2}\right)} 
\int\!d\tau\!\int\!d\tau'\,
\frac{\Phi_0^I(\tau)\Phi_0^I(\tau')}{(\tau-\tau')^{2\Delta}}\,. 
\label{value}
\end{eqnarray}

\medskip  

In the next we will discuss a holographic relation between (\ref{value}) and 
a source term insertion on a straight Wilson line.

\subsection{GKPW-type relation}
\label{hol}

Let us argue the role played by the quadratic action $S_{(2)}$ 
in the dual gauge-theory side from now on. For simplicity we restrict 
ourselves to the bosonic part again. 

\medskip 

We begin our argument with the holographic relation {\it without} the
quadratic fluctuations. That is, we focus upon $S_{(0)}$\,. For
the straight Wilson line
\[
W \equiv \Tr P {\rm e}^{\int\!dt\,\left(iA_0 + \phi_6\right)}\,,
\]
the following relation is well known: 
\begin{eqnarray} 
\langle W \rangle = {\rm e}^{-S_{(0)}} =1\,. \label{Wst}
\end{eqnarray}
Note that $S_{(0)}=0$ for the straight line. 

\medskip 

The GKPW relation says that the right-hand side of (\ref{Wst})
is the leading term of the semiclassical string partition function
around the static AdS$_2$ solution corresponding to the straight Wilson
line. Thus we see that
\begin{eqnarray}
Z_{\rm string} =  
\int [d\Phi]~ {\rm e}^{-S_{\rm full}[\Phi]}
\approx  {\rm e}^{-S_{(0)}} =1\,.  \label{partition-fn}
\end{eqnarray}
Here $S_{(0)}$ has been evaluated by putting $\Phi=\Phi_{\rm cl}$ into
the action $S_{\rm full}[\Phi]$\,, where $\Phi_{\rm cl}$ describes the
classical solution corresponding to the 1/2 BPS Wilson line.

\medskip 

On the other hand, considering the semiclassical expansion around the
classical solution in the string side, the Wilson loop corresponding to
the string classical solution is inserted into the partition function of
$\mathcal{N}=4$ SYM,
\begin{eqnarray}
 Z_{\rm SYM} = \int\! [dA][d\phi]\,{\rm e}^{-S_{\rm SYM}[A,\phi]} 
 \longrightarrow 
 \int\! [dA][d\phi]\, W\, {\rm e}^{-S_{\rm SYM}[A,\phi]} 
 = \langle W \rangle\,.
\end{eqnarray}
Thus (\ref{Wst}) gives a piece of evidence for the conjectured relation
\[
Z_{\rm SYM} = Z_{\rm string}\,,
\]
at the leading order of the semiclassical approximation. 

\medskip 

Then the next problem to be considered is to add the quadratic action
$S_{(2)}$ to (\ref{Wst}).  It is an easy task to add the contribution of
$S_{(2)}$ to (\ref{partition-fn}). The field $\Phi$ should be decomposed
into the classical solution and the fluctuation like $\Phi = \Phi_{\rm
cl} + \Phi_{\rm fl}$\,, and the full action $S_{\rm full}$ is expanded
with respect to the fluctuations $\Phi_{\rm fl}$\,. The resulting
expression is
\begin{eqnarray}
Z_{\rm string} \approx 
{\rm e}^{-S_{(0)}} \int [d {\Phi}_\mathrm{fl}]~
{\rm e}^{-S_{(2)} [\Phi_{\rm fl}]}
= \int [d {\Phi}_\mathrm{fl}]~
{\rm e}^{-S_{(2)} [\Phi_{\rm fl}]}\,. 
\label{quad}
\end{eqnarray}
Thus the string partition function is approximated by a path integral in
terms of the fluctuation. 

\medskip 

Now we shall discuss a further semiclassical approximation to
(\ref{quad})\,. The quadratic action $S_{(2)}$ describes a collection of
free theories on AdS$_2$\,. Then the GKPW relation may be applied to the
world-sheet theory on the AdS$_2$ like in a scalar field theory on
AdS$_5$\,.

\medskip 

It would be helpful to remember some arguments for the Lorentzian
AdS/CFT correspondence \cite{BKL}. In the Lorentzian AdS/CFT, $\Phi_{\rm
fl}$ has to be regarded as the sum of NN mode $\Phi_{\rm
NN}$ and normalizable mode $\Phi_{\rm N}$ as follows: \cite{BKL}
\[
 \Phi_{\rm fl} = \Phi_{\rm NN} + \Phi_{\rm N}\,.
\]
Here $\Phi_{\rm NN}$ is characterized by the boundary value
$\Phi_0$\,. Then the NN and normalizable modes should be
treated as the background and fluctuation, respectively. 
Note that the result of \cite{DGT} is reproduced by setting that
$\Phi_{\rm NN}=0$\,.

\medskip 

In our context this should be the so-called ``second semiclassical
approximation" or ``double holography."  It is not so surprising to see
the holography of this type since other examples are already known in
the case of defect CFT \cite{DFO}.  Observing that the AdS$_2$ solution
is a kind of probe inserted in the bulk, such as probe D-branes in the
bulk AdS$_5$\,, it is quite natural to see the holography even for the
present case.

\medskip 

There however exists no normalizable solution (i.e., $\Phi_{\rm N}=0$)
in Euclidean signature. Thus we propose the following holographic
relation 
\begin{eqnarray}
\left\langle \Tr P\bigl[ {\rm e}^{\int\!dt\,\left(iA_0 + \phi_6\right)} 
\cdot {\rm e}^{\int\!dt\,\mathcal{O}_I\cdot \Phi_0^I} \bigr]
\right\rangle =  {\rm e}^{-S_{(2)}[\Phi_0]}
\,. \label{proposal}
\end{eqnarray}
The $S_{(2)}[\Phi_0]$ is obtained by putting the NN mode
$\Phi_{\rm fl}=\Phi_{\rm NN}$ into the $S_{(2)}[\Phi_{\rm fl}]$ and it
is given by (\ref{value}). 
Note that 
the boundary values couple to the operators
in one dimension
rather than four dimensions. The other three coordinates are fixed as
the same as the position of the Wilson line.

\medskip 

By taking the functional derivatives in terms of the sources
$\Phi_0$'s and then setting that $\Phi_0=0$\,, it is possible to 
produce correlation functions of the operators $\mathcal{O}$'s
inserted on the one-dimensional Wilson line. The gravity side is a
collection of the NN modes of the scalar fields on the
AdS$_2$ and so it is natural to argue that the correlation functions
should be related to the CQM via
AdS$_2$/CFT$_1$\,.

\subsection{The coupled operators}
\label{4.5}

So far we have not specified what operators should be inserted as
$\mathcal{O}_I$\,. The coupled operators have already been clarified in
our previous works \cite{SY:NR2,SY:NR3}.

\medskip 

\begin{figure}
 \begin{center}
  \includegraphics[scale=.7]{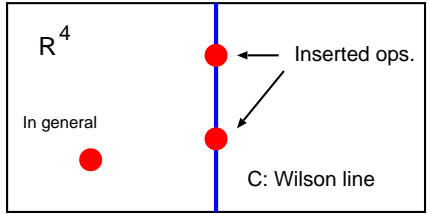}
 \end{center}
\caption{\footnotesize The restriction of the insertion points of the
 operators onto the Wilson line $C$\,. In general the operators may be
 inserted everywhere on $\mathbb{R}^4$\,.}\label{insertion:fig} 
\end{figure}

The coupled operators $\mathcal{O}$ are 
\begin{eqnarray}
\mathcal{O}_{I} = \left\{\begin{array}{cc}
iF_{a0} + D_{a}\phi_6 & \quad  (a=1,2,3) \\ 
\phi^{a'} & \quad (a'=1,\ldots 5)
\end{array}
\right.\,.  \nn
\end{eqnarray}
The operators with $\Delta =2$ and $\Delta=1$ correspond to massive and
massless scalars, respectively. This is well supported from the GKPW
relation between mass of scalar fields and conformal dimensions for
AdS$_2$ \cite{AdS/CFT:GKPW}\,,
\begin{eqnarray}
\Delta = \frac{1}{2}\left(1+\sqrt{1+4R_0^2 m_{\rm B}^2}
\right)\,. \label{GKPW}
\end{eqnarray}
With this relation (\ref{GKPW}) we obtain that 
\[
\Delta =2 \quad \mbox{for} \quad m^2_{\rm B}=2/R_0^2\,, \qquad 
\Delta =1 \quad \mbox{for} \quad m^2_{\rm B}=0\,.
\]
The GKPW relation has been used for the AdS$_2$ case above. One can
figure out this as a restriction of the positions of the operators
inserted at the boundary. Taking AdS$_2$ rather than AdS$_5$\,, the
operators are inserted just on the Wilson line rather than everywhere in
$\mathbb{R}^4$\,. For an intuitive picture see
Fig.\,\ref{insertion:fig}. So far the bosonic insertion has been
considered but the fermionic insertion also can be discussed. The
fermionic insertion is discussed in \cite{SY:NR3} and it is $\Psi$
satisfying $P_+\Psi = \Psi$\,. Thus it has the eight physical components
and its conformal dimension is 3/2.

\medskip 

From another viewpoint, the source term insertion may be regarded as a
deformation of the straight Wilson line, which was discussed in
\cite{SY:NR2} by following \cite{Miwa}. In our previous works
\cite{SY:NR2,SY:NR3} the physical meaning of the small fluctuation
$\delta C$ around the straight Wilson line $C_0$ was unclear, but now
one can realize that $\delta C$ should be given by adding the source
terms. In fact, the Wilson loop expansion discussed in \cite{SY:NR2} can
be reproduced by expanding the source term up to the contact terms (For
the second order expansion of the Wilson line see Appendix
\ref{2nd}). For the first order it can easily be checked by identifying
$\Phi_0$ with the deformations $\delta x^{a}$ and $\delta \dot  y^{a'}$\,.

\medskip 

In the string-theory side 16 linear supersymmetries are possessed by the
quadratic action $S_{(2)}$\,, so the same amount of supersymmetries
should be preserved even after the source term  has been
inserted. This is shown in Appendix \ref{susy}.

\subsection{A comment on higher-order fluctuations}

Finally let us comment on higher-order fluctuations, e.g., $S_{(3)}$\,.

\medskip 

\begin{figure}
\begin{center}
\includegraphics[scale=.7]{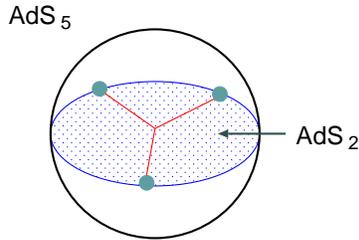}
\end{center}
\caption{\footnotesize A contribution of the third-order fluctuation}
\label{int:fig}
\end{figure}

In the Euclidean case the fluctuations around the classical solution
corresponding to the Wilson line are nothing but the NN
modes. Then higher order fluctuations depend on the boundary value $\Phi_0$
like $S_{(i)}[\Phi_0]$\,. For example, a contribution of the third-order
fluctuation $S_{(3)}[\Phi_0]$ gives a triple coupling on the AdS$_2$
world-sheet as depicted in Fig.\ \ref{int:fig}.

\medskip 

The higher-order fluctuations are not discussed in this paper, but it
would be interesting to consider them carefully as a future direction.

\section{Conclusion and discussion}

We have discussed a holographic dual of the NR string on \adss{5}{5}.
The physical spectrum of the NR string \cite{ST} completely agrees with
that of a one-dimensional CQM. Then the wave functions and the two-point
function of the CQM can be reproduced from the physical modes of the NR
string.  From this result we have argued that an AdS$_2$/CFT$_1$ may be
realized in AdS$_5$/CFT$_4$\,. In Euclidean signature there exist only
NN solutions for the equation of motion of the NR
string. Then the boundary values related to a source term giving the
operator insertions on the Wilson line. The source term may be regarded
as a deformation of the Wilson loop. Thus we have proposed a GKPW-type
relation between the NR string and the deformed Wilson line by the
source term insertion.

\medskip 

There remains an open problem that is to derive the CQM action directly
from $\mathcal{N}$=4 SYM or the D-brane setup. The NR limit is closely
related to the Higgs mechanism in $\mathcal{N}$=4 SYM. This would be
obvious by considering the D3-brane setup before taking the near-horizon
limit. The fluctuations around the AdS$_2$ solution correspond to those
around an infinitely long open string. Thus the CQM may describe a
quantum mechanics of the probe quarks supplied from the long string. It
would be interesting to study the origin of the CQM in this direction.

\medskip 

It would also be nice to consider a further generalization of our
arguments. The first example is to discuss a circular case. The
quadratic action has already been discussed in \cite{DGT} and the
corresponding Wilson loop expansion has been done in \cite{SY:NR3}. The
remaining problem is to derive normalizable modes on AdS$_2$ by using
the Poincare disk, instead of the strip considered in \cite{ST}. It may
be expected that the dual CQM is defined on the circle given by the
Wilson loop.

\medskip

The second example is to consider higher-dimensional representation
instead of the fundamental representation. Then the Wilson loop should
be replaced by giant Wilson loops \cite{gWilson} and we have to use
D-brane actions on AdS$_5\times$S$^5$ instead of the string action. From
the viewpoint of the gauge theory the operator insertion should not be
modified. The only difference is the representation of the trace. Thus
the quadratic fluctuations around the giant Wilson loops should behave
as those around AdS$_2$ in the vicinity of the boundary. Thus we can
easily guess the agreement. We will report on the detail computation in
the near future \cite{future}.

\medskip 

We hope that our research may be able to shed light on AdS$_2$/CFT$_1$
correspondence.

\section*{Acknowledgment}

The authors would like to thank N.~Iizuka, A.~Jevicki, Y.~Kazama,
A.~Miwa, Y.~Okawa, T.~Okuda, T.~Takayanagi, D.~Trancanelli, A.~Tsuji,
N.~Yokoi and T.~Yoneya. They also thank the YITP workshop YITP-W-07-05
on ``String Theory and Quantum Field Theory'' and the international
workshop ``Progress of String Theory and Quantum Field Theory''
(2007/12/7-10, Osaka City University). Discussions there 
were useful to complete this work.

\medskip 

The work of M.S.\ was supported in part by the Grant-in-Aid for
Scientific Research (19540324,\,19540304,\,19540098) from the Ministry
of Education, Science and Culture, Japan. The work of K.Y.\ was
supported in part by JSPS Postdoctoral Fellowships for Research Abroad
and the National Science Foundation under Grant No.\,NSF PHY05-51164.

\appendix

\section*{Appendix}

\section{NR limit in another coordinate system of AdS$_5\times$S$^5$}
\label{nr}

The NR limit given in \cite{GGK} is written with a complicated, 
unfamiliar metric. Here we shall rewrite the limit in terms of more 
familiar coordinate system. 

\medskip 

The following coordinates for AdS$_5\times$S$^5$\,, which keeps an
AdS$_2$ factor explicitly, would be useful:
\begin{eqnarray}
ds^2 &=& R_0^2\bigl[
\cosh^2u\, (-\cosh^2 x\, dt^2 + dx^2) + du^2 
+ \sinh^2 u\, (d\varphi_1^2 + \sin^2\phi_1 d\phi_2^2 ) \bigr] \nonumber \\
&& + R_0^2\bigl[d\gamma^2 + \cos^2\gamma\, d\varphi_3^2 
+ \sin^2\gamma \left(d\psi^2 + \cos^2\psi\, d\varphi_1^2
+ \sin^2\psi\, d\varphi_2^2\right) 
\bigr]\,.
\end{eqnarray}
Here the S$^5$ part is also represented by 
\begin{eqnarray}
&& X_1+iX_2 = \sin\gamma \cos\psi {\rm e}^{i\varphi_1}\,, \qquad 
X_3+iX_4 = \sin\gamma \sin\psi {\rm e}^{i\varphi_2}\,, \nonumber \\ 
&& X_5+iX_6 = \cos\gamma {\rm e}^{i\varphi_3}\,, \qquad \qquad 
X_1^2 + \cdots + X_6^2 = 1\,. 
\end{eqnarray}

\medskip 

Then the NR limit in \cite{GGK} can be taken as follows. 
\begin{eqnarray}
 u=\frac{\tilde{u}}{R_0}\,, \qquad \gamma = \frac{r}{R_0}\,,
  \qquad 
\varphi_3 =\frac{\pi}{2} 
+ \frac{y}{R_0}\,, \qquad R_0 \rightarrow \infty\,.
\label{limit}
\end{eqnarray}
This limit keeps the AdS$_2$ factor in AdS$_5$ while the geometry
around a point
is closed up. Thus it respects an SO(3)$\times$SO(5) symmetry preserved
by the corresponding Wilson line.

\medskip

The resulting metric is given by 
\begin{eqnarray}
ds^2 &=& (R_0^2 + \tilde{u}^2)(-\cosh^2 x\, dt^2 + dx^2) + d\tilde{u}^2 
+ \tilde{u}^2\, (d\phi_1^2 + \sin^2\phi_1 d\phi_2^2 ) \nonumber \\ 
&& + dr^2 + dy^2 
+ r^2 \left(d\psi^2 + \cos^2\psi\, d\varphi_1^2
+ \sin^2\psi\, d\varphi_2^2\right)\,, \label{nr1} 
\end{eqnarray}
which is equipped with the divergent NS-NS two-form flux
\begin{eqnarray}
 B_{tx} = R_0^2 \cosh x\,. \label{nr2}
\end{eqnarray}
It basically describes the geometry of AdS$_2\times \mathbb{R}^8$\,, though the
leading term is divergent. This is nothing but the metric obtained after
taking the NR limit in \cite{GGK}.

\section{A heuristic interpretation of the NR string action}
\label{der}

The original derivation of the NR string is complicated and it seems
difficult to figure out the essential point. So we will not repeat it
but give a heuristic interpretation of the NR string action. In
particular we aim at understanding how the mass terms can appear in the
quadratic action. It would be important to have a rough image for the
mechanism to generate the mass terms.

\medskip 

We start from the background (\ref{nr1}) with (\ref{nr2})\,, which was
obtained from AdS$_5\times$S$^5$ background with a constant NS-NS
two-form flux after taking the NR limit (\ref{limit})\,. For simplicity
we restrict ourselves to the bosonic part.

\medskip 

The Nambu-Goto action on this background is given by 
\begin{eqnarray}
S^{({\rm NR})} &=& -\frac{1}{4\pi\alpha'}
\int\!d^2\sigma\,
\sqrt{-\det g}
+
\frac{1}{2\pi\alpha'}\int\!d^2\sigma\,B_{tx}  
\,, \nonumber 
\end{eqnarray}
where the induced metric is given by 
\begin{eqnarray}
&&  g_{ij} = (R_0^2+(x^a)^2))g_{0ij} + \partial_i x^a \partial_j x^a 
+ \partial_i y^{a'} \partial_j y^{a'}\,. \nonumber \\ 
&& g_{0ij} = {\rm diag}(-\cosh^2\sigma, 1)\,, \nonumber 
\end{eqnarray}
by using the static gauge 
\[
 t=\tau\,, \qquad x = \sigma\,. 
\]
Here the coordinates have been rewritten as 
\begin{eqnarray}
&& d\tilde{u}^2 
+ \tilde{u}^2\, (d\phi_1^2 + \sin^2\phi_1 d\phi_2^2 ) = 
\sum_{a=1}^3dx^{a}dx^a\,, \nonumber \\ 
&& dr^2 + dy^2 
+ r^2 \left(d\psi^2 + \cos^2\psi\, d\varphi_1^2
+ \sin^2\psi\, d\varphi_2^2\right) 
= \sum_{a'=1}^5dy^{a'}dy^{a'}\,. \nonumber 
\end{eqnarray}

\medskip 

Then the action can be expanded as 
\begin{eqnarray}
S^{\rm (NR)} = -\frac{1}{4\pi\alpha'}\int\!d^2\sigma
\sqrt{-\det g_{0}}\,g_0^{ij}\left(
\partial_i x^a \partial_j x^a + g_{0ij}(x^a)^2 + 
\partial_i y^{a'} \partial_j y^{a'}
\right) + \mathcal{O}\left(R_0^{-2}\right)\,. \nonumber
\end{eqnarray}
The divergence in $R_0\to\infty$ limit has been canceled out due to the
presence of the NS-NS two-form (\ref{nr2})\,. The higher-order terms
also disappear in this limit. It is important to observe that the value
of the mass 2 counts the number of the world-sheet coordinates, namely
$\tau$ and $\sigma$ (or the AdS$_2$ factor in the metric
(\ref{nr1})). It would be easy to extend the above argument to NR
D-brane cases \cite{SY:NR1,SY:NR2}.

\medskip

It is also helpful to remember the pp-wave case, where the mass term
comes from the (++)-component of the metric, $ds^2= -2dx^+ dx^- +
G_{++}(dx^{+})^2 +\cdots$\,. In analogy to the pp-wave, $(dx^+)^2$
correspond to the AdS$_2$ factor in the present case.

\medskip 

Thus we have understood how the mass terms should come up in the action
and the meaning of the value of the mass.

\section{Some properties of Gegenbauer polynomial}
\label{gegen}

Here we shall summarize some useful properties of Gegenbauer polynomial
in investigating normalizable modes of a scalar field on AdS$_2$\,.

\medskip

The Gegenbauer polynomial is represented by the hyper geometric function
or Jacobi polynomial $P_{\alpha}^{(a,b)}(z)$ with
$a=b=\lambda-1/2$
\begin{eqnarray}
C_{\alpha}^{\lambda}(z) 
&=& \frac{\Gamma(\alpha+2\lambda)}{\Gamma(\alpha+1)\Gamma(2\lambda)} 
F(\alpha+2\lambda,-\alpha,\lambda+\frac{1}{2}; \frac{1-z}{2}) \nn \\ 
&=& \frac{\Gamma(\lambda+\frac{1}{2})}{\Gamma(2\lambda)}
\frac{\Gamma(\alpha+2\lambda)}{\Gamma(\alpha+\lambda+\frac{1}{2})}
P_{\alpha}^{(\lambda-1/2,\lambda-1/2)}(z)\,. \nn 
\end{eqnarray}
It satisfies the following hyper-geometric differential equation 
\begin{eqnarray}
\left[(1-z^2)\frac{d^2 }{dz^2} - (2\lambda +1)\frac{d}{dz} 
+ \alpha(\alpha+2\lambda) 
\right] C_{\alpha}^{\lambda} =0\,. \nn
\end{eqnarray}

\medskip 

When $\lambda$ is real and $\lambda > -1/2$\,, the Gegenbauer polynomials
satisfy the orthonormal condition given by
\begin{eqnarray}
\int^1_{-1}\!\!dx\,(1-x^2)^{\lambda-1/2}C_{m}^{\lambda}(x)
C_{n}^{\lambda}(x) = \frac{\pi\Gamma(n+2\lambda)}
{2^{2\lambda-1}n! (\lambda+n)\Gamma(\lambda)^2}
\delta_{m,n}\,. \nn 
\end{eqnarray}

\medskip 

The Gegenbauer polynomials are also given by the generating function
\[
 \frac{1}{(1-2z t +t^2)^{\lambda}}
=\sum_{\alpha=0}^{\infty}C_{\alpha}^{\lambda}(z)t^{\alpha}\,.
\]
The first few Gegenbauer polynomials are 
\begin{eqnarray}
&& C_{0}^{\lambda}(z)=1\,, \qquad C_1^{\lambda}(z) = 2\lambda z\,, 
\qquad C_{2}^{\lambda}(z)=-\lambda +2\lambda(1+\lambda)z^2\,, \nn \\ 
&& C_{3}^{\lambda}(z)=-2\lambda(1+\lambda)z
 +\frac{4}{3}\lambda(1+\lambda)(2+\lambda)z^3\,.\nn 
\end{eqnarray}

\section{Wilson loop expansion at the second order}
\label{2nd}

Let us discuss a Wilson loop expansion at the second order 
with respect to a small deformation of the loop. 

\medskip 

A Wilson loop with a contour $C$ is described by 
\begin{eqnarray}
W(C)=\Tr \CW_{u_1}^{u_2}~,~~~
\CW_{u_1}^{u_2} = P \exp \int_{u_1}^{u_2} \!\!ds\, \left(
iA_\mu(x(s))\dot x^\mu(s)+\phi_i(x(s))\dot y^i
\right)\,. \nn
\end{eqnarray}
The locally supersymmetry condition
\begin{eqnarray}
(\dot x^\mu)^2-(\dot y^i)^2=0  
\label{LSC}
\end{eqnarray}
can be viewed as the integrability condition 
of the super-invariance of $W(C)$\,:
\[(i\Gamma_\mu\dot x^\mu+\Gamma_i\dot y^i)\epsilon=0\,.
\]

\medskip 

Let us consider a small deformation of $W(C)$ by taking $C$ as 
$C = C_0+\delta C$: 
\[
x^\mu=x^\mu_{C_0}+\delta x^\mu\,, \qquad y^i=y_{C_0}^i+\delta y^i\,.
\]
Then $W(C)$ can be expanded as
\begin{eqnarray}
W(C)&=&W(C_0)
+\int_{u_1}^{u_2}\!\!ds\, 
\Bigg[
\delta x^\mu(s)\frac{\delta W(C)}{\delta x^\mu(s)}\Big|_{C_0}
+
\delta \dot y^i(s)\frac{\delta W(C)}{\delta \dot y^i(s)}\Big|_{C_0}
\Bigg]
\cr&&
+\int_{u_1}^{u_2}\!\!ds_1\int_{u_1}^{u_2}\!\!ds_2\,
\Bigg[
\delta x^\mu(s_1)\delta x^\nu(s_2)
 \frac{\delta^2 W(C)}{\delta x^\mu(s_1)\delta x^\nu(s_2)}\Big|_{C_0}
+\cdots
\Bigg]+\cdots\,. \nn
\end{eqnarray}
It is straightforward to derive\footnote{We denote $iA_\mu(x(s))$ as $(iA_\mu)_s$ for short.}
\begin{eqnarray}
\frac{\delta\CW_{u_1}^{u_2}}{\delta x^\mu(s)}&=&
\CW_{u_1}^s
O_\mu(s)
\CW_s^{u_2}
+\CW_{u_1}^{u_2}(iA_\mu)_s\delta(u_2-s)
-(iA_\mu)_s\delta(u_1-s)\CW_{u_1}^{u_2}~, 
\nn\\
\frac{\delta\CW_{u_1}^{u_2}}{\delta \dot y^i(s)}&=&
\CW_{u_1}^sO_i(s)\CW_s^{u_2}\,.
\label{kihon}
\end{eqnarray}
Here we have introduced $O_{\mu}$ and $O_{i}$ defined as, respectively, 
\begin{eqnarray}
O_\mu \equiv iF_{\mu\nu}\dot x^\nu
+D_\mu\phi_i \dot y^i\,, \qquad 
O_i \equiv \phi_i\,, \nn
\end{eqnarray}
where 
$F_{\mu\nu}=\partial_\mu A_\nu -\partial_\nu A_\mu
+i[A_\mu,A_\nu]~$ and $D_\mu\phi_i=\partial_\mu\phi_i+i[A_\mu,\phi_i]~$.

\medskip 

In the first equation in (\ref{kihon}), we have performed a
partial-integration and used the following properties:
\begin{eqnarray}
\partial_u\CW_{u_1}^u&=&\CW_{u_1}^u
(iA_\mu\dot x^\mu+\phi_i\dot y^i)_u\,, \quad
\partial_u\CW_{u}^{u_2}=
-(iA_\mu\dot x^\mu+\phi_i\dot y^i)_u
\CW_{u}^{u_2}\,.
\label{hashi}
\end{eqnarray}
From (\ref{kihon}) one can read off the operator insertion preserving
half of the supersymmetries. This is analogy with the pp-wave case
\cite{Miwa}.

\medskip

A 1/2 BPS straight Wilson line is realized by choosing $C_0$ as 
\begin{eqnarray}
C_0&:&
x^\mu=(s,0,0,0)\,, \qquad 
\dot y^i=(0,0,0,0,0,1)\,, \nn
\end{eqnarray}
which satisfies the locally supersymmetry condition (\ref{LSC}). Then 
$W(C_0)$ represents a straight line with $u_1=-\infty$ and $u_2=+\infty$\,, 
which is represented by 
\begin{eqnarray}
W(C_0)=\Tr \CW_0{}_{u_1}^{u_2}\,, \qquad 
\CW_0{}_{u_1}^{u_2}=\CW_{u_1}^{u_2}|_{C_0}=
P\exp\int_{u_1}^{u_2}du (iA_0+\phi_6)_u\,. \nn 
\end{eqnarray}

\medskip 

From (\ref{kihon}), the first order deformation of $W(C_0)$
is given by 
\begin{eqnarray}
\frac{\delta W(C)}{\delta x^\mu(s)}\Big|_{C_0}&=&
\Tr~  \CW_0{}_{u_1}^{s}\CO_{\mu}(s)
\CW_0{}_s^{u_2}\,, \nn  \\
\frac{\delta W(C)}{\delta\dot y^i(s)}\Big|_{C_0}&=&
\Tr~ \CW_0{}_{u_1}^{s}\CO_i(s)
\CW_0{}_s^{u_2}\,, \nn
\end{eqnarray}
where 
\begin{eqnarray}
\CO_\mu=O_\mu|_{C_0}=iF_{\mu0}+D_\mu\phi_6\,, \qquad 
\CO_i=O_i|_{C_0}=\phi_i\,. \nn
\end{eqnarray}
Then the second-order derivatives are 
\begin{eqnarray}
\frac{\delta^2 W(C)}{\delta x^\mu(s_1)\delta x^\nu(s_2)}\Big|_{C_0}
&=&\Tr \Big[
\CW_0{}_{u_1}^{s_1}\CO_\mu(s_1)\CW_0{}_{s_1}^{s_2}\CO_\nu(s_2)
\CW_0{}_{s_2}^{u_2}
\cr&&~~~
+\CW_0{}_{u_1}^{s_1}
\big(iD_{(\mu} F_{\nu)0}
+D_{(\mu} D_{\nu)}\phi_6
\big)
\CW_0{}_{s_1}^{u_2}\delta(s_2-s_1)
\cr&&~~~
-\frac{1}{2}\CW_0{}_{u_1}^{s_2}
iF_{\mu\nu}(s_2)
\partial_{s_2} \delta(s_2-s_1)
\CW_0{}_{s_2}^{u_2}
\cr&&~~~
+\frac{1}{2}\CW_0{}_{u_1}^{s_1}
iF_{\mu\nu}(s_1)
\partial_{s_1} \delta(s_2-s_1)
\CW_0{}_{s_1}^{u_2}
\Big]
~,\nn\\
\frac{\delta^2 W(C)}{\delta \dot y^i(s_1)\delta x^\mu(s_2)}\Big|_{C_0}&=&
\Tr \Big[
\CW_0{}_{u_1}^{s_1}
\CO_i(s_1)
\CW_0{}_{s_1}^{s_2}\CO_\mu(s_2)\CW_0{}_{s_2}^{u_2}
\cr&&~~~
+\CW_0{}_{u_1}^{s_2}(D_\mu\phi_i)_{s_2}\delta (s_2-s_1) \CW_0{}_{s_2}^{u_2}
\Big]~,~~~\nn\\
\frac{\delta^2 W(C)}{\delta x^\mu (s_1)\delta \dot y^i(s_2)}\Big|_{C_0}&=&
\Tr \Big[
\CW_0{}_{u_1}^{s_1}\CO_\mu(s_1)
\CW_0{}_{s_1}^{s_2} \CO_i(s_2)
\CW_0{}_{s_2}^{u_2}
\cr&&~~~
+\CW_0{}_{u_1}^{s_2}(D_\mu\phi_i)_{s_2}\delta (s_2-s_1) \CW_0{}_{s_2}^{u_2}
\Big]~,~~~\nn\\
\frac{\delta^2 W(C)}{\delta \dot y^i(s_1)\delta \dot y^j(s_2)}\Big|_{C_0}&=&
\Tr \Big[
\CW_0{}_{u_1}^{s_1}
\CO_i(s_1)
\CW_0{}_{s_1}^{s_2}
\CO_j(s_2)
\CW_0{}_{s_2}^{u_2}
\Big]\,, \nn
\end{eqnarray}
where  $s_1\le  s_2$ is assumed.
We have used (\ref{kihon}) and (\ref{hashi}),
and performed partial integrations.
It is straightforward to derive higher order deformations, 
but we will not touch on them here.

\medskip 

Note that the above second-order deformations contain the contact terms which look
 like Schwinger terms. Now we are not sure for the physical
interpretation of these terms in our context. The Wilson loop expansion
coincides with the Wilson loop with the source term
\begin{eqnarray}
\Tr P\left[
 \e^{\int dt (iA_0+\phi_6)}
\,\e^{\int dt \CO_I\Phi^I}
\right]\,,
\label{WL with O}
\end{eqnarray}
up to
these contact terms.

\section{Supersymmetry of deformed Wilson loop}
\label{susy}

We expect 16 supersymmetries preserved by the deformed Wilson line
$W(C)$ from the result in the string-theory side.  The purpose here is
to show that it is really supersymmetric.

\medskip  

First of all, let us recall the relation between world-sheet
fluctuations and NN modes.
As seen in section 4.3, the world-sheet fluctuations
\begin{eqnarray}
(\sqrt{2\pi}\lambda^{-1/4}\tilde x^a, \sqrt{2\pi}\lambda^{-1/4}\tilde
y^{a'}) \nonumber 
\end{eqnarray}
 are small comparing to the classical contribution as
$\lambda\to \infty$\,.  

\medskip 

The NN modes are characterized by the behavior in the vicinity of the
boundary as \eqref{NN boundary behavior}.
As $(x,y)$ and $(\tilde x,\tilde y)$ are related by \eqref{nn},
we find 
\[
\tilde x \to x_0(\tau)\,, \qquad \tilde y \to y_0(\tau)\,, 
\] 
as approaching the boundary. This implies that $(\tilde x,\tilde y)$ do
not grow large near the boundary. Thus the world-sheet fluctuations near
the boundary
\[
 (\sqrt{2\pi}\lambda^{-1/4} x_0^a, \sqrt{2\pi}\lambda^{-1/4}
 y^{a'}_0)\,
\]
still remain small. Those may cause small deformations of the Wilson
line. 

\medskip 

Now we shall consider supersymmetries preserved by $W(C)$\,. The
supersymmetry condition is given by the locally supersymmetry condition
\eqref{LSC} imposed on the coordinates at a point on $C$\,. For the
first-order deformation, the condition reads \cite{SY:NR2} 
\begin{eqnarray*}
(\dot x^\mu_{C_0} + \delta\dot x^\mu)^2-
(\dot y^\mu_{C_0} + \delta\dot y^\mu)^2=
2(\delta\dot x^0-\delta\dot y^6)=0\,. 
\end{eqnarray*}
Since $\delta\dot x^0+\delta\dot y^6=0$ can be obtained by using SO(2)
symmetry, we may choose 
\begin{equation}
\delta x^0=\delta\dot y^6=0\,. \label{su-cond}
\end{equation}

\medskip 

Next it is the turn to consider the second-order deformation at a point.
It corresponds to the terms with a $\delta$-function in the Wilson line
expansion and
this should not be confused with two first-order deformations at two
different points on $C$\,.  Let us denote the second-order deformation
as $(\delta^2 x^\mu, \delta^2 y^{i})$\,. Then 
it is obvious that the condition is trivially
satisfied as follows:
\begin{eqnarray*}
(\dot x^\mu_{C_0} + \delta\dot x^\mu+ \delta^2\dot x^\mu)^2-
(\dot y^\mu_{C_0} + \delta\dot y^\mu+ \delta^2\dot y^\mu)^2=
2(\delta\dot x^0-\delta\dot y^6)=0\,,
\end{eqnarray*}
because the second-order deformation is small comparing to the
first-order deformation. 

\medskip 

This is the case for higher-order deformations at a point and $W(C)$
coincides with (\ref{WL with O}) up to contact terms.  Thus we have
shown that $W(C)$ is supersymmetric under the condition (\ref{su-cond}).
Inversely speaking, the consistency with the supersymmetries requires the
condition (\ref{su-cond}).

\end{document}